\documentclass[a4paper,11pt]{article}

\usepackage[english]{babel}
\usepackage[utf8x]{inputenc}
\usepackage[T1]{fontenc}
\usepackage{multirow}
\usepackage[a4paper,top=2cm,bottom=2cm,left=2cm,right=2cm,marginparwidth=1.75cm]{geometry}
\usepackage{color}
\usepackage{soul}
\usepackage[numbers, sort&compress]{natbib}
\usepackage{amsmath}
\usepackage{bm}
\usepackage{caption}
\usepackage{subcaption}
\usepackage{graphicx}
\usepackage{url} 
\usepackage{authblk}
\title{Run-away transition to turbulent strong-field dynamo}
\author[1,2,*]{A.~Guseva}
\author[1]{L.~Petitdemange}
\author[2]{S.~M.~Tobias}
\affil[1]{LERMA, l'Observatoire de Paris, Sorbonne Université, Université  PSL, CNRS, 75014 Paris, France}
\affil[2]{Department of Applied Mathematics,  University of Leeds, Leeds LS2 9JT, UK }
\affil[*]{Corresponding author: anna.guseva@obspm.fr}
\date{}

\newcommand{\anna}[1]{{ #1}}

\graphicspath{{./}{images/}}

\begin{document}

\maketitle

\begin{abstract}

Planets and stars are able to generate coherent large-scale magnetic fields by helical convective motions in their interiors.  This process, known as hydromagnetic dynamo, involves nonlinear interaction between the flow and the magnetic field. Nonlinearity facilitates existence of bi-stable dynamo branches: a weak field branch where the magnetic field is not strong enough to enter into the leading order force balance in the momentum equation at large flow scales, and a strong field branch where the field enters into this balance. The transition between the two with enhancement of convection can be either subcritical or supercritical, depending on the strength of magnetic induction. In both cases, it is accompanied by topological changes in velocity field across the system; however, it is yet unclear how these changes are produced. In this work, we analyse transitions between the weak and strong dynamo regimes using a data-driven approach, separating different physical effects induced by dynamically active flow scales. Using Dynamic Mode Decomposition, we decompose the dynamo data from direct numerical simulations into different components (modes), identify the ones relevant for transition, and estimate relative magnitudes of their contributions Lorentz force and induction term.  Our results suggest that subcritical transition to a strong dynamo is facilitated by a subharmonic instability, allowing for a more efficient mode of convection, and provide a modal basis for reduced-order models of this transition.
\end{abstract}

\section*{Plain Language Summary}
Planetary magnetic fields are generated by stretching, twisting, and folding of magnetic field lines by flow motions inside planetary cores. This process is called planetary dynamo; it is driven by convection between the hotter inner core of a planet and its outer layers. As magnetic fields get stronger, they exert a magnetic tension force on these motions and modify them. In this work, we study how the dynamo can transition from a ``weak" state, where this feedback is negligible, to a ``strong" state, where magnetic tension becomes comparable to other dominant forces affecting the flow, buoyancy and Coriolis. We analyse the data from several numerical models of this process and find that it takes place through a succession of periodic flow states that are unstable and thus observed during only a limited period of time. The presence of these state allows the flow to transport heat more efficiently, and therefore generate stronger magnetic fields. Transitions as those described here can help to understand formation and evolution of planetary magnetic fields.

\section{Introduction}\label{sec:intro}
The flow dynamics in planetary cores can be greatly affected by large-scale magnetic fields,  generated by helical convective motions in a process known as hydromagnetic dynamo \cite{moffatt1978magnetic}. In the context of stability theory, dynamo may arise as a kinematic instability of  non-magnetic convection when magnetic induction is larger than magnetic diffusion so that a certain ratio between them, critical magnetic Reynolds number $Rm$, is achieved. \anna{This condition is typically fulfilled for planetary flows; e.g. $Rm\sim O(10^3)$ for geodynamo~\cite{schaeffer2017turbulent}.} 
An observable  manifestation of planetary dynamos are magnetic torsional waves  detected in secular variations  of the Earth magnetic field \cite{gillet2010fast,finlay2023gyres}  and Jupiter luminosity \cite{hori2023jupiter}. They arise when magnetic tension enters in the dominant balance with buoyancy (Archimedean) and the Coriolis force in the momentum equation, so-called MAC \anna{or magnetostrophic} balance \cite{dumberry2003torque}.  This force balance, corresponding to a ``strong"-field dynamo, constrains magnetic field to the Taylor state. In this state, the azimuthal component of magnetic tension, averaged over cylindrical surfaces parallel to the rotation axis \anna{of a planet}, tends to zero,  because all other forces in azimuthal direction are either zero or negligible \cite{taylor1963magneto}. Considerable progress to obtain planetary-like Taylor state dynamos was made through analysing mean-field dynamo \anna{equations}, driven by a parametrized electromotive force \cite{fearn2004evolution} and using adjoint-based optimization \cite{li2018taylor}.

At the same time, self-consistent numerical models of planetary dynamos, driven solely by convection, were able to reproduce predominantly dipolar magnetic field topology both in Boussinesq and anelastic models \cite{glatzmaiers1995three, jones2014dynamo}. However, realistic planetary parameter regimes are computationally unachievable, because the rotation time scale is much faster compared to the time scale of  viscous or magnetic dissipation. 
The ratio between the two time scales, Ekman number, is estimated to be $Ek=10^{-14}$ for the Earth, $10^{-16}$ for Jupiter, and $10^{-17}$ for Saturn \cite{gastine2014explaining}, in contrast to $Ek = 10^{-7}-10^{-6}$ accessible to the state-of-art direct numerical simulations \anna{\cite{schaeffer2017turbulent,aubert2019approaching}}. It was proposed to reduce \anna{simultaneously} the Ekman number and the magnetic Prandtl number $Pm$, the ratio between viscosity and magnetic diffusivity, \anna{in order to} keep the ratio of time scales in simulations as close as possible to the Earth-like regime \cite{davidson2013scaling,aubert2017spherical, yadav2016approaching,schaeffer2017turbulent}. Even then, simulations result in the dominant quasi-geostrophic (QG) balance between the Coriolis force and the pressure gradient at the largest flow scales, reproducing  MAC balance only at the second order, after the geostrophic, gradient components of forces are removed \cite{aubert2017spherical,teed2023solenoidal}.  The flows in QG force balance, aligned along the rotation axis due to the Taylor-Proudman theorem, can also support steady large-scale fields, although in a ``weak"-field regime where the magnetic Lorentz torque is balanced by viscous force \cite{hollerbach1996theory}. Such ``weak"-field dynamos are less likely to excite magnetic waves observed in planets because the restoring magnetic force would be too weak.

Another strategy to reach MAC regime is to decrease magnetic diffusivity as compared to viscosity while keeping $Ek$ relatively high, i.e. to increase the magnetic Prandtl number \cite{dormy2016strong,petitdemange2018systematic}. Weaker magnetic dissipation allows to obtain stronger magnetic fields and stronger Lorentz force. This approach works well close enough to the onset of convection; \anna{away from it, enhanced} convective forcing also increases inertia and may lead to dipole breakdown and multipolar magnetic fields \cite{christensen2006scaling}, unless the field is strong enough to extend the stability domain of dipolar dynamos~\cite{menu2020magnetic}. 
However, it is computationally less expensive and allows to explore a range of different parameter regimes of ``weak" and ``strong"-field dynamos, and transitions between them. At relatively low $Pm$, these transitions are gradual, or supercritical: mildly chaotic, multi-modal turbulent convection excites weak axisymmetric dynamo mode that gradually builds up its energy in the saturated state with enhancement of \anna{convection} \cite{petitdemange2018systematic}. Through nonlinear Lorentz force, it reduces the level of turbulence in the flow or even suppresses it to laminar, one-mode convection. On the other hand, at large $Pm$, a ``weak" dynamo is easily excited by near-critical convectively unstable flow modes, even when convection is not turbulent. In this case, an abrupt, subcritical transition to turbulent yet predominantly dipolar ``strong"-field dynamo can take place when convective forcing is only slightly increased. Such transitions are essentially nonlinear processes of interaction between the field and the flow through the Lorentz force and induction term, and  can be relevant for developing dynamos in young planets or failed dynamos in formerly magnetized planets like Mars \cite{hori2013subcritical}. It is yet not well understood how the corresponding topological changes in magnetic and velocity fields affect the dynamo action, and what triggers turbulence in the case of abrupt, subcritical transitions at high $Pm$.

In this work, we aim to fill this gap through analysis of subcritical transition from the weak to the strong dynamo, combining direct numerical simulations (DNS) and a data-driven approach (section~\ref{sec:eqn}). In complex three-dimensional  dynamos, a large number of degrees of freedom is required to represent the flow and the field in DNS.  We employ a data-driven analysis method, Dynamic Mode Decomposition (DMD), to represent physical processes behind the transition in a more compact way and to identify principal dynamical components in the data from dynamo simulations, as described in section~\ref{sec:data}. DMD seeks for the closest linear approximation to a nonlinear system \cite{schmid2022dynamic}, and was recently used to identify small- and large-scale periodic dynamics in one-dimensional dynamo models~\cite{guseva2024data} and quasi-Keplerian  magnetohydrodynamic turbulence~\cite{guseva2023transition}. In section~\ref{sec:sw} we decompose the dynamo flow into convective Rossby wave and its subharmonic, appearing during transition. Section~\ref{sec:forces} describes how nonlinear interactions between the flow and the field are affected by presence of these structures. Finally, we analyse the breakdown of the subharmonic mode to a chaotic state  through a secondary instability (section~\ref{sec:runaway}) and discuss our results in section~\ref{sec:discussion}.

\section{Dynamo equations and methods}\label{sec:eqn}


%
\begin{figure}
\centering
\includegraphics[width=0.4\textwidth]{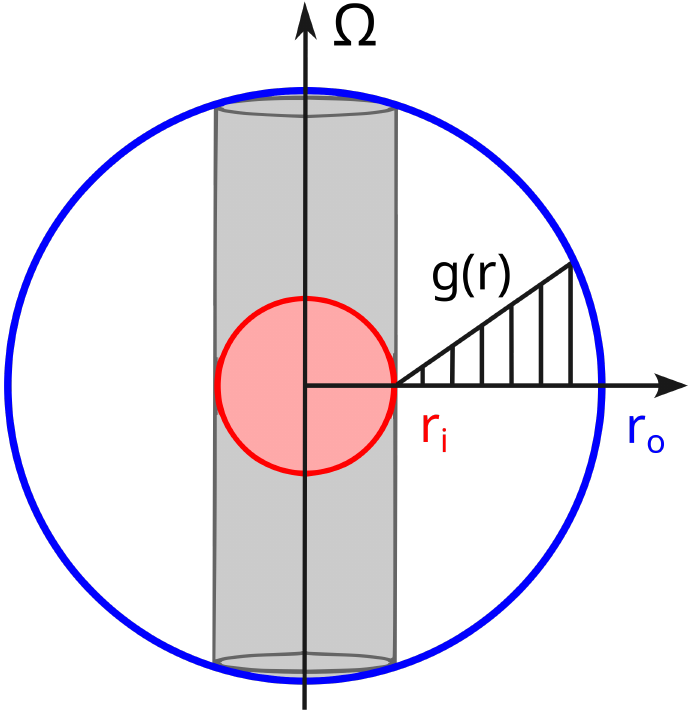}      
\caption{The computational domain. Conducting fluid is confined between two co-rotating spheres and Boussinesq approximation is assumed, with gravity $g$ linearly proportional to the spherical radius. The grey area indicates the tangent cylinder, the imaginary surface circumscribing the inner sphere.}
\label{fig:setup}
\end{figure}

We solve coupled equations for velocity $\mathbf{u}$, magnetic field $\mathbf{B}$ and temperature $T$ for incompressible fluid in  the Boussinesq approximation in a rotating spherical shell, 
\begin{align}
   & E(\frac{\partial \mathbf{u}}{\partial t } + \mathbf{u} \cdot \nabla \mathbf{u} \anna{- \nabla^2 \mathbf{u}}) = -\nabla P -2 \mathbf{z} \times \mathbf{u} + Ra \frac{\mathbf{r}}{r_0} T + \frac{1}{Pm} (\nabla \times \mathbf{B}) \times \mathbf{B} \nonumber \\
   & \frac{\partial \mathbf{B}}{\partial t }  = \nabla \times(\mathbf{u} \times \mathbf{B}) + \frac{1}{Pm} \nabla^2 \mathbf{B} \label{eq:NStInd} \\
 &\frac{\partial T}{\partial t} + \mathbf{u} \cdot \nabla T = \frac{1}{Pr} \nabla^2 T, \nonumber \\
 &\nabla \cdot \mathbf{B} =0, \anna{\quad \nabla \cdot \mathbf{u}=0}. \nonumber
\end{align}
Equations~\eqref{eq:NStInd} are solved in spherical coordinates $(r, \theta,\phi)$; we also introduce cylindrical coordinate system $(s, \phi, z)$, aligned with the axis of rotation, for data analysis. Convection is driven by an imposed temperature difference $\Delta T = T_i - T_o$ between the cooler outer sphere of radius $r_o$ and the hotter inner sphere of radius $r_i = 0.35 r_o$. Figure~\ref{fig:setup} illustrates the computational setup of two concentric spheres, rotating with constant angular velocity $\Omega$ and containing conducting fluid with density $\rho$, viscosity $\nu$, thermal diffusivity $\kappa$ and magnetic diffusivity $\eta$. The grey area denotes the tangent cylinder, an imaginary cylindrical surface around the inner sphere with cylindrical radius $s = r_i$.   The equations are non-dimensionalized with $\Delta T$, the gap width between the spheres $d=r_o-r_i$, viscous time $d^2/\nu$,  and the magnetic scale $(\rho \mu \eta \Omega)^{1/2}$. The Ekman number was set to $Ek = \nu/\Omega d^2 =10^{-4}$, the magnetic Prandtl number to $Pm = \nu/\eta = 12$, and the thermal Prandtl number to $Pr=\nu/\kappa =1$.  The modified Rayleigh number $Ra = \alpha g_0 \Delta T d /\nu \Omega$, where $\alpha$ is thermal expansivity and $g_0$ is
gravitational acceleration at $r_o$, controls convection strength. $Ra$ was varied in the parameter regime where transitions between weak- and strong-field dynamo topologies were previously observed \cite{petitdemange2018systematic}. Equations~\eqref{eq:NStInd} were solved numerically with pseudospectral DNS code PaRoDy \cite{dormy1998mhd}, with no-slip boundary conditions for velocity and insulating boundary conditions for the magnetic field on both spheres.  In  spectral space, the flow variables were represented by spherical harmonics with azimuthal periodicity, or wavenumber, $m$ and spherical degree $l$.  We used  $N_r =240$ radial points and  $N_\theta = 72$, $N_\phi = 2 N_\theta$ spherical harmonics with $3/2$ dealiasing rule.

 Dynamic Mode Decomposition  is a data-driven approach for identification of coherent structures in dynamical systems like~\eqref{eq:NStInd}  \cite{schmid2022dynamic}. Nonlinear magnetohydrodynamic equations~\eqref{eq:NStInd} evolve state vector $\mathbf{q}(t) = (\mathbf{u},\mathbf{B},T)$ of the flow variables at a given point $\bm{\mu}$ in the parameter space, 
\begin{equation}\label{eq:nonl_sys}
    \frac{d \mathbf{q}}{d  t} = f^{\mathrm{nonlinear}} (\mathbf{q},t,\bm{\mu}), \quad \bm{\mu} = (Ra, E, Pm, Pr).
\end{equation}
According to Koopman theory~\cite{koopman1931hamiltonian}, a generic nonlinear system~\eqref{eq:nonl_sys} can be equivalently represented with a linear operator $\mathcal{K}$, propagating in time an infinite number of flow observables $f(\mathbf{q})$ as  $f (\mathbf{q}) (t_{k+1}) = \mathcal{K} f (\mathbf{q}) (t_{k})$ at each moment in time $t_{k}$. The eigenfunctions and eigenvalues of $\mathcal{K}$ correspond thus to the dominant flow components and their evolution in time. Dynamic Mode Decomposition approximates the Koopman operator using matrices of DNS data, 
\begin{equation}\label{eq:dmd}
    Q = \left[q_1, q_2, \cdots q_{k-1}\right], \quad Q' = \left[q_2, q_3, \cdots q_{k}\right], \quad Q' = \mathcal{A} Q 
\end{equation}
 collected at  times $t_1, t_2, \cdots t_k$ \cite{tu2014dynamic,arbabi2017ergodic}. In essence, matrix $ \mathcal{A}$ is a data-driven linearization of the system~\eqref{eq:NStInd} about its nonlinear state, obtained from simulations or measurements. For computational stability, it is practical to get first its reduced representation $A_r$ in the space of principal orthogonal modes $\Phi_r$ of the data, using singular value decomposition of the data matrix $Q = \Phi_r \Sigma_r V_r^*$ and further linear matrix transformations, 
 \begin{equation}\label{eq:Ar}
  Q' \approx  \mathcal{A} \Phi_r \Sigma_r V_r^*, \qquad \Phi_r^* \mathcal{A} \Phi_r = \Phi_r^* Q' V_r \Sigma^{-1}_r = \mathcal{A}_r,
  \end{equation}
\anna{where $^*$ denotes conjugate transpose of a matrix.} The elements of matrices $\Sigma$ and $V$ correspond to the energy and temporal evolution of \anna{the orthonormal basis} $\Phi$, respectively. The number of retained modes of this basis, model rank $r$, is a parameter of the DMD approximation, setting the number of linearly independent components. In this work it was set to retain $99\%$ of the energy in each dynamo variable in \eqref{eq:NStInd}. \anna{The eigenfunctions $\tilde{\psi}$ and the discrete-time eigenvalues $\lambda$ of $\mathcal{A}_r$ represent the dynamics of  system~\eqref{eq:nonl_sys} in reduced $r$-dimensional space. DMD modes $\psi$ in physical space and DMD eigenvalues $\omega$ in continuous time representation are reconstructed from them according to}
  \begin{equation}\label{eq:DMD_linsys}
\mathcal{A}_r \tilde{\psi} = \lambda \tilde{\psi}, \qquad \psi = \frac{1}{\lambda} Q' V_r \Sigma^{-1}_r \tilde{\psi}, \qquad \omega = \ln(\lambda)/ \Delta t.
 \end{equation}
The real and imaginary parts of $\omega$, $\Re(\omega)$ and $\Im(\omega)$, correspond to modal growth rates and frequencies, respectively. See \citet{schmid2010dynamic,tu2014dynamic,arbabi2017ergodic,schmid2022dynamic} for more details on the DMD algorithm and the underlying Koopman theory.

\section{Transition between weak and strong dynamos in simulations}\label{sec:data}
\begin{figure}
     \centering
\includegraphics[width=\textwidth]{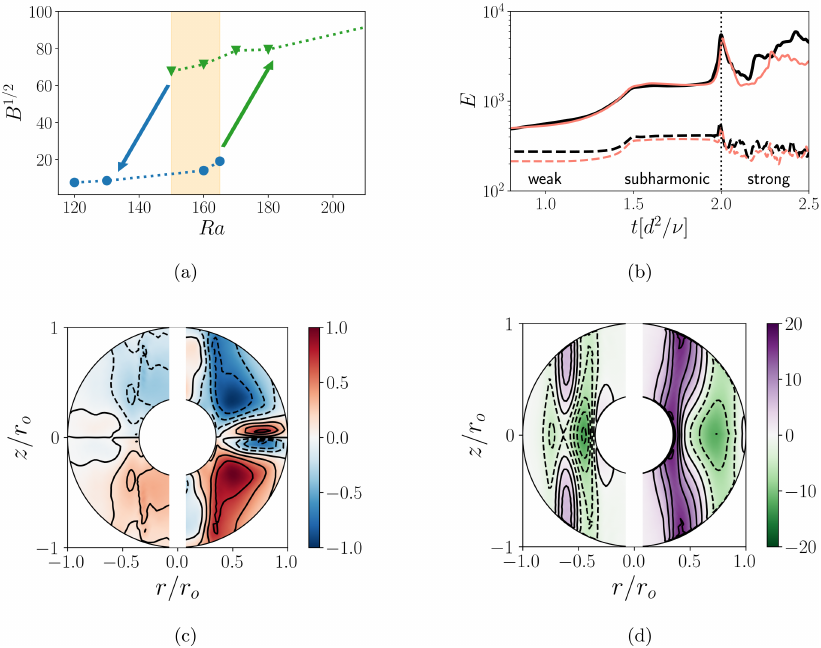}      
        \caption{(a) Rms amplitude of magnetic field of the weak (blue circles) and strong dynamos (green triangles); shaded region denotes co-existence of the two branches. Green arrow indicates transition from the weak to strong branch (panel b) and the blue arrow inverse transition (figure~\ref{fig:chaos_m}b). 
 (b) Kinetic (dashed) and magnetic volume-integrated energy (solid) during transition. In black: $Ra=180$, initial condition of weak dynamo at $Ra=165$. In red: $Ra=170$, initial condition at $Ra=160$. Time series are shifted in time for comparison so that transition to strong-field dynamo takes place at $t=2$ (dotted line).  (c) Radial component of axisymmetric magnetic field $B_r$, averaged over $\phi$ and time, in the weak (left) and the strong-field regime (right). (d) The same as (c) but  for the axisymmetric zonal flow \anna{(equation \ref{eq:zonal_flow})}. In (c) and (d), colours denote the mean radial magnetic field and  the zonal flow, \anna{respectively}, and contours highlight the change of sign and extrema of these quantities. } 
        \label{fig:w2s_dns}
\end{figure}

\anna{To track both weak and strong dynamo branches, we ran initial
simulations with small random perturbations in the fields at $Ra=160$ to find weak-field states. These weak-field states were then used as initial conditions for simulations at higher $Ra$, which either remained on the weak-field branch or transitioned to the
strong-field branch. The strong-field solutions were then used as the starting point for further simulations, tracking flow states in figure~\ref{fig:w2s_dns}a. The three different types of initial conditions are summarized in table~\ref{tab:simulations}.}

In the first, the flow was initialized with small random perturbations of magnetic, velocity and temperature fields. Such initial conditions result first in fast exponential growth of convective modes, followed by much slower growth of the magnetic field (see section~\ref{sec:discussion} for more details). Eventually, the rms amplitude of the magnetic field saturates into a statistically steady state on the weak-field branch (figure~\ref{fig:w2s_dns}a). The second type of initial condition is this weak-field dynamo solution at lower $Ra$; the simulations are initialized with weak-field magnetic, velocity and temperature fields, and $Ra$ is \anna{increased} in the simulation parameters and the simulation is run until a steady-state is obtained again.  
\anna{The solutions of these runs stay in the weak-field branch for $Ra < 165$, and run away to the strong-field branch when $Ra > 165$ (figure~\ref{fig:w2s_dns}a).} A typical dynamo evolution during such simulations is depicted in figure~\ref{fig:w2s_dns}(b), with kinetic and magnetic energy during such transitions for $Ra=170$ and $Ra=180$, and two different weak-field initial conditions at $Ra=160$ and $Ra=165$, respectively. The energies are integrated over all spherical harmonics and radius. Initially, the system settles down into a weak-field solution  ($t<1.3$); it is characterized by predominantly dipolar axisymmetric field inside the tangent cylinder  perturbed by the convective thermal Rossby wave near equator (figure~\ref{fig:w2s_dns}c, left). In this regime, the flow consists of a single convective mode and is constrained vertically through Taylor-Proudman theorem. The axisymmetric zonal flow,
\begin{equation}\label{eq:zonal_flow}
  V_\phi = \langle u_\phi \rangle_\phi,  
\end{equation}
 where \anna{$ \langle \cdots \rangle_\phi$} denotes an average over $\phi$, is weak in this regime. It is formed by  retrograde jets near both the tangent cylinder and the equator (figure~\ref{fig:w2s_dns}d, left), separated by prograde flow regions. This initial \anna{weak-field} state is unstable, and both magnetic and kinetic energies gradually increase when $t\in [1.3, 1.5]$, saturating into another quasi-steady solution. This solution, active when $t \in [1.5, 2.0]$, is also unstable, and after a rapid run-away growth in both kinetic and magnetic energies, the flow transitions to a chaotic dynamo state driven by turbulent convection, i.e. the strong dynamo branch in figure~\ref{fig:w2s_dns}(a). The magnetic field is still predominantly dipolar, with a strong axisymmetric $l=1$ component (figure~\ref{fig:w2s_dns}c, right), yet now this field is predominantly outside of the tangent cylinder. In this regime, convection is multi-modal, kinetic energy is much smaller than magnetic (figure~\ref{fig:w2s_dns}b), and a strong prograde zonal flow develops (figure~\ref{fig:w2s_dns}d, right) along the tangent cylinder. The larger $Ra$, the less time the flow spends in the vicinity of the weak-field state, however, the overall time and evolution path taken by this transition remain independent of $Ra$, or different weak-field initial conditions \anna{(figure~\ref{fig:w2s_dns}b)}. 
 
 States that result from such transitions as the one described above are known to exhibit hysteresis, or bi-stability. We use the third type of initial conditions, magnetic, velocity and temperature snapshots from the strong-field branch, to initialize dynamo simulations at lower $Ra$, essentially tracking back this branch until it ceases to exist. When a dynamo simulation is initialized with a strong-field chaotic state, it remains on the strong-field dynamo branch as $Ra$ is gradually reduced. Only when convective forcing decreases to  $Ra\leq 150$, the flow gradually relaxes into a weak-field non-turbulent state, losing energy in chaotic components and saturating to the dominant convective mode. Thus, in the bi-stability region of $Ra\in [150, 165]$, denoted in figure~\ref{fig:w2s_dns}(a) by the shaded area, both strong and weak-field solutions coexist. Finally, the energy for the run with $Ra=120$ is slowly decaying, so the dynamo potentially \anna{ceases to operate when} $Ra\leq120$.

\begin{table}[h!]
\centering
 \begin{tabular}{|p{7cm} |c |c |c|} 
 \hline

 Initial condition (IC) & $Ra$ of IC & Simulation $Ra$  & Final state\\
 \hline 
Small random perturbations in magnetic field, temperature and velocity &  &  160  & Weak-field\\ 
 \hline
 \multirow{ 4}{7cm}{Fully developed convective dynamo on the weak-field branch} & 160   & 165 & Weak-field\\ 
  & 160 & 170  & Strong-field\\
  & 160 & 180  & Strong-field\\
   &  165  & 180 & Strong-field\\
 \hline
  \multirow{ 4}{7cm}{Fully developed convective dynamo on the strong-field branch}   & 180 & 350  & Strong-field\\ 
  & 170  & 160  & Strong-field\\
  & 160 & 150  & Strong-field\\
   & 160 & 140  & Weak-field\\
   & 150  & 130 & Weak-field\\
   & 150  & 120  & Weak-field\\
 \hline
 \end{tabular}
 \caption{Summary of the simulations performed in this work, with their initial conditions and final states. These simulations were used to map the dynamo bifurcation diagram in figure~\ref{fig:w2s_dns}a. The rest of parameters were set to $Ek=10^{-4}$, $Pm=12$ and $Pr=1$. $Ra$ number does not vary in time during simulations.}
\label{tab:simulations}
\end{table}

In this work, we will  analyse in detail three different simulations with similar transition scenarios, the first one at $Ra=170$ and initial conditions at $Ra=160$, the second one at $Ra=180$, with the same initial conditions, and the third one again at $Ra=180$ but with the initial state at $Ra=165$. \anna{All of them were initialized with a snapshot of the saturated weak-field dynamo.} 
We collect flow data for these three cases during the time interval in figure~\ref{fig:w2s_dns}(b), and analyse separately four stages of transition: the weak dynamo, the transient high-energy state, the  run-away increase in dynamo energy, and the relaxation to the chaotic strong-field dynamo.

\section{DMD modes and their evolution during transition}
\begin{figure}
     \centering
\includegraphics[width=\textwidth]{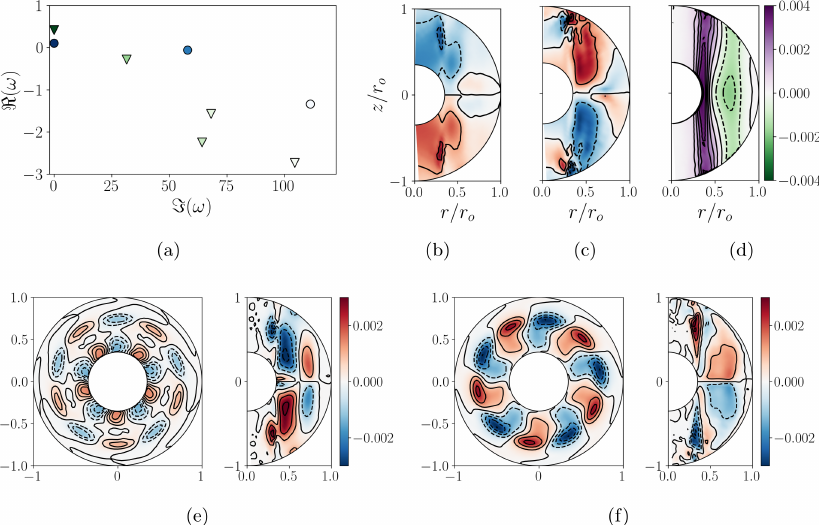}      
        \caption{\anna{Dynamic Mode Decomposition of dynamo \anna{data} at $Ra=180$, with initial conditions at $Ra=160$.} (a) DMD spectrum of temperature $T$; $\Re(\omega)$ and $\Im(\omega)$ denote the growth rates and frequencies of the dynamo calculated according to~(\ref{eq:DMD_linsys}).  Blue circles, the ``weak" phase;  green triangles, transition to strong dynamo regime. \anna{Symbol shading corresponds to the magnitude of DMD modes, calculated from the best fit of the data to the modal basis $\psi$~\eqref{eq:DMD_linsys}. (b) The  ``mean" mode of weak magnetic field $B_r$, with frequency $\Im(\omega)=0$. (c) Same mode but during transition. (d)  The ``mean" mode of azimuthal velocity component $u_\phi$, of the transient subharmonic state. (e) Equatorial and latitudinal cross-sections of the convective Rossby wave in $B_r$, $\Im(\omega)=60$. (f) Subharmonic mode of $B_r$ with $\Im(\omega)=30$, identified during transition. In panels (b-f), contours are used to highlight the change of sign and extrema in the DMD modes. }}
        \label{fig:dmd_modes}
\end{figure}

We perform DMD of different components of DNS dynamo data in the weak ($t<1.3$) and transient regimes ($1.3<t<2.0$) separately, using the algorithm described in section~\ref{sec:eqn}. Since temperature, magnetic and velocity fields are related through nonlinear terms in~\eqref{eq:NStInd}, all the flow variables contain some contribution from the dominant flow components due to nonlinear quadratic interactions in the flow. We thus show the typical spectrum of DMD eigenvalues for  temperature $T$ in figure~\ref{fig:dmd_modes}(a), colored by their best-fit magnitudes \cite{jovanovic2014sparsity}. During the weak phase, the dynamo has only two principal modes: a convective Rossby wave, resulting from vertical alignment of convective cells due to rotation, and the mean flow modes with $\Im(\omega) =0$. The Rossby wave has a frequency of $\Im(\omega) \approx 60$ for $Ra=180$ and $\Im(\omega) \approx 80$ for $Ra=170$, and comprises five pairs of convective cells in azimuthal direction, corresponding to the azimuthal wavenumber $m=5$. The decrease in frequency with $Ra$ is expected as nonlinear effects can slow down prograde rotation of convective eigenmodes \cite{feudel2013multistability,skene2024weakly}. Figure~\ref{fig:dmd_modes}(c) illustrates the spatial shape of this mode for the radial component of magnetic field, antisymmetric with respect to the equator. Convective cells of this structure are inclined and rotate in the direction of global rotation. Rossby waves, driven by convective temperature gradient, generate mean dipolar magnetic field component in $B_r$, $B_\theta$, and $B_\phi$. This is identified by DMD as the mean-field mode with frequency of $\Im(\omega)=0$, which has the strongest amplitude in the magnetic field both in weak and transient regimes. Its spatial structure \anna{(figure~\ref{fig:dmd_modes}b)} correlates well with the mean field (figure~\ref{fig:w2s_dns}c, left). In temperature, convective mixing \anna{induced} by the Rossby wave also results in the mean mode with $\Im(\omega)=0$ (figure~\ref{fig:dmd_modes}a), corresponding to deformation and  flattening of initially linear mean temperature profile. The $\Im(\omega)=0$ contributions to $u_r$ and $u_\theta$, i.e. meridional flows, are absent or very weak. The $\Im(\omega)=0$ component of $u_\phi$, corresponding to zonal flow $V_\phi$,  represents less than $10\%$ of the flow energy in the weak-field regime, very small amount compared to the Rossby waves.
\label{sec:sw}

In the transient subharmonic regime, the Rossby wave becomes less coherent and is now identified by DMD as damped (figure~\ref{fig:dmd_modes}a); instead, a subharmonic mode of the Rossby wave overtakes the dynamics. It also has azimuthal periodicity of $m=5$ but half the frequency, $\Im(\omega)=30$, and thus two times slower rotation in the azimuthal direction. Figure~\ref{fig:dmd_modes}(d) shows a meridional slice of this mode for $B_r$, where the outer convective cells have expanded over the equatorial area.  In the vertical direction, a local maximum appeared at the tangent cylinder, where the mode changes sign. The contours of this structure are  less aligned with the direction of rotation. The subharmonic component drives a strong zonal flow \anna{(figure~\ref{fig:dmd_modes}d)}, detectable in the DMD spectrum of $u_\phi$ (not shown here). This zonal flow is  inverted in comparison to the weak state (figure~\ref{fig:w2s_dns}d, left) and take form of strong prograde jet \anna{(purple-colored)} near the tangent cylinder and weaker retrograde jet \anna{(green-colored)} at the equator. Development of the zonal flow is accompanied by topological changes in magnetic field. Strong $V_\phi$ promotes toroidal field $\langle B_\phi \rangle_\phi$; to  satisfy divergence-free condition, axisymmetric $\langle B_r \rangle_\phi$ and $ \langle B_\theta \rangle_\phi$ also adjust.  The shape of  $\Im(\omega)=0$ mode of magnetic field changes considerably \anna{(figure~\ref{fig:dmd_modes}c)}, so that  the local maxima in $B_\phi$ move from the regions inside the tangent cylinder towards its boundary. Magnetic field becomes concentrated at the tangent cylinder, and the regions inside it become stagnant and contain less magnetic flux than in the weak-field state. 

The remaining component of the DMD spectrum in figure~\ref{fig:dmd_modes}(a) with higher frequency $\Im(\omega)>100$ corresponds to quadratic interaction of the Rossby wave with itself and thus has azimuthal periodicity of $m=10$, yet preserving its vertical structure. When the subharmonic is dominant, similar modes corresponding to quadratic nonlinear interactions of subharmonic flow components also appear (see an additional $\Im(\omega)\approx 60$ mode in the transient dynamo spectrum in  figure~\ref{fig:dmd_modes}a). \anna{These quadratic components are nevertheless much weaker than the dominant modes, if compared by magnitude.}

\begin{figure}
    \centering
\includegraphics[width=\textwidth]{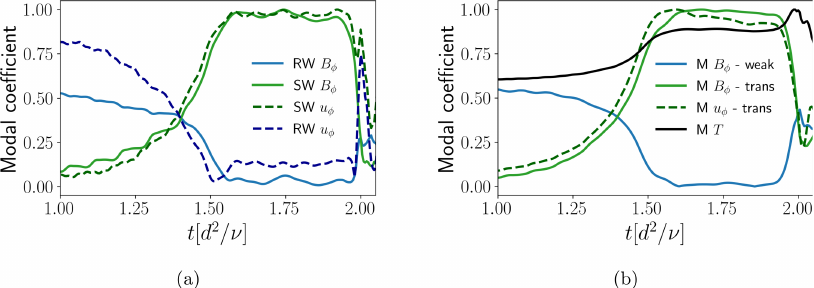}      
    \caption{Absolute values of modal coefficients. (a) In blue, the Rossby wave (RW) obtained in the weak regime, in green, its subharmonic (SW). Solid lines, $B_\phi$; dashed, $u_\phi$.  (b) Coefficients of the mean (M) modes  with $\Im(\omega)=0$. Solid blue (green), from $B_\phi$ in the weak (transient) regime.  Dashed green, from $u_\phi$ in the transient subharmonic state. In black, the coefficient of mean-field mode of temperature, which is similar for the weak and the subharmonic states. $Ra=170$, initial conditions at $Ra=160$.}
    \label{fig:dmd_coeffs}
\end{figure}

 Further insight on the dynamical behaviour of the mean modes, Rossby and subharmonic,  can be obtained by projecting the DNS dynamo data onto the identified above modal basis, using oblique least-squares projection (see~\citet{guseva2024data} for more details). This procedure gives  instantaneous coefficients of the modes, and allows to follow their temporal evolution during all four phases of transition.  Figure~\ref{fig:dmd_coeffs}(a) shows \anna{that} the absolute values of the Rossby wave  coefficient  decrease both in $B_\phi$ and $u_\phi$, while the amplitude of the subharmonic grows and saturates for a limited period of time in the transient state, with energy exchange between the two during this transition. The Rossby wave loses its coherence, and becomes nearly extinct in components of magnetic field, but retains some energy in the flow. The amplitude of the mean  mode of temperature grows by a factor of $1.5$, with further flattening of mean temperature profile in the bulk of the flow and more efficient heat transport (figure~\ref{fig:dmd_coeffs}b). The zonal flow mode grows nearly by $90$\%, correlated with the growth of the mean toroidal field $B_\phi$. Since the zonal flow is weak in the weak dynamo state (figures~\ref{fig:w2s_dns}d,f), we can ignore spatial reconfiguration of $V_\phi$ as the subharmonic grows. It is partially accounted for by the  change of sign of the modal coefficient of the corresponding subharmonic mode, when the flow in the weak-field regime is projected on it.

\section{Evolution of forces during transition}\label{sec:forces}
\subsection{Modal contribution to induction and Lorentz forces}

\begin{figure}
    \centering
\includegraphics[width=\textwidth]{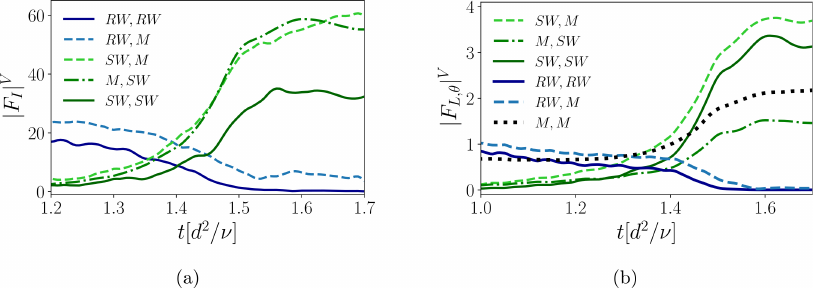}      
    \caption{(a) Modal contributions to the integrated induction term $|F_I|^V$~\eqref{eq:ind_force_int}.  Solid, interaction of $\mathbf{u}^{RW(SW)}$ with $\mathbf{B}^{RW(SW)}$; dashed, $\mathbf{u}^{RW (SW)}$ with $\mathbf{B}^{M}$; dashdotted, $\mathbf{u}^{M}$ with $\mathbf{B}^{SW}$. Blue color corresponds to the Rossby wave and green color to the subharmonic. (b) \anna{Modal contributions to the volume-integrated $\theta$-component of the Lorentz force~\eqref{eq:lorentz_force_int}.  Solid, interaction of current $\nabla \times \mathbf{B}^{RW(SW)}$ with the field component $\mathbf{B}^{RW(SW)}$; dashed, $\nabla \times \mathbf{B}^{RW (SW)}$ with $\mathbf{B}^{M}$; dashdotted, $\nabla \times \mathbf{B}^{M}$ with $\mathbf{B}^{SW}$.} The dotted line represents interaction of $\mathbf{B}^{M}$ with the current induced by it, $\nabla \times \mathbf{B}^{M}$. $M$ denotes the mean mode, $RW$ denotes the Rossby wave, and $SW$ its subharmonic.   $Ra=180$, initial conditions at $Ra=165$. }
    \label{fig:dmd_forces}
\end{figure}

In this section, we estimate the contribution of the Rossby and subharmonic modes to the induction term and Lorentz force from equations~\eqref{eq:NStInd},
\begin{equation}\label{eq:FI_FL}
\mathbf{F}_I = \nabla \times (\mathbf{u} \times \mathbf{B}), \qquad \mathbf{F}_L = \frac{1}{Pm} (\nabla \times \mathbf{B}) \times \mathbf{B},
\end{equation}
responsible for nonlinear interactions between the flow and the field, and leading to the dynamo saturation.
 Reconstructing velocity and magnetic fields from the modes (figures~\ref{fig:dmd_modes}b-f) and their temporal coefficients (figure~\ref{fig:dmd_coeffs}), we decompose velocity and magnetic fields as
\begin{equation}\label{eq:field_decomp}
        \mathbf{u} = \mathbf{u}^M + \mathbf{u}^{RW} + \mathbf{u}^{SW} + \cdots, \quad \mathbf{B} = \mathbf{B}^M + \mathbf{B}^{RW} + \mathbf{B}^{SW} + \cdots 
\end{equation}
Here $M$ denotes the mean mode, which is equivalent to the zonal flow $V_\phi$ in velocity field, $RW$ denotes the Rossby wave, and $SW$ its subharmonic. The corresponding expansion for induction term,
\begin{align}
   \mathbf{F}_I &= \nabla \times (\mathbf{u}^M \times \mathbf{B}^M) + \nabla \times (\mathbf{u}^{RW} \times \mathbf{B}^{RW})  + \nabla \times (\mathbf{u}^{SW} \times \mathbf{B}^{SW}) \nonumber \\
   &+ \nabla \times (\mathbf{u}^M \times \mathbf{B}^{RW})  + \nabla  \times (\mathbf{u}^{RW} \times \mathbf{B}^M) 
+ \nabla \times (\mathbf{u}^M \times \mathbf{B}^{SW})  \label{eq:force_decomp}\\ 
&+ \nabla \times (\mathbf{u}^{SW} \times \mathbf{B}^M) + \nabla \times (\mathbf{u}^{SW} \times \mathbf{B}^{RW}) + \nabla \times (\mathbf{u}^{RW} \times \mathbf{B}^{SW}) + \cdots \nonumber, 
\end{align}
isolates interaction of different velocity and magnetic structures: mean, Rossby and subharmonic waves with themselves, interaction of the mean with the Rossby or the subharmonic mode, and interaction of the Rossby wave and its subharmonic.  Such interactions are in essence triadic interactions between two waves exciting a third one which is allowed if their frequencies fulfill the condition $\omega_3 = \omega_1 \pm \omega_2$, and their wavenumbers match $m_3 = m_1 \pm m_2$  (see  e.g.  \citet{barik2018triadic} for a discussion on triadic interactions in spherical shells).

Figure~\ref{fig:dmd_forces}(a) shows the rms value of the strongest terms from expansion~\eqref{eq:force_decomp} to the total induction, integrated over the computational domain,
\begin{equation}\label{eq:ind_force_int}
   |F_{I}|^V = \left[(1/V)\int_V (F_{I,r}^2 + F_{I,\theta}^2 + F_{I,\phi}^2) dV \right]^{1/2},
\end{equation} 
as a function of time. During the weak dynamo phase, induction has the largest contribution from the interaction of the Rossby wave with the mean magnetic field \anna{(dashed blue line)}, adding a perturbation to magnetic field with the same azimuthal periodicity of $m=5$ and the frequency of the Rossby wave through triadic interaction $(\textbf{u}^{RW}, \textbf{B}^{M}, \textbf{B}^{RW})$ in the quadratic terms~\eqref{eq:FI_FL}. The second largest component is the interaction  of this perturbation  with the Rossby wave in the flow \anna{(solid blue line)}, feeding back into the mean dipole component of magnetic field through $(\textbf{u}^{RW}, \textbf{B}^{RW}, \textbf{B}^{M})$ or its second harmonic with $\Im(\omega) = 2 \Im(\omega)_{RW}$ flow component, also detected by DMD (figure~\ref{fig:dmd_modes}a). This interaction can be interpreted as $\alpha$-effect,  an electromotive force generating large-scale magnetic fields  from correlated fluctuations of the flow and the field in the framework of mean-field dynamo theory~\cite{moffatt1978magnetic,moffatt2019self}. Since it appears in all components of induction, and the zonal flows are weak, the weak-field state is essentially an $\alpha^2$-dynamo \cite{schrinner2007mean,schrinner2012dipole}. 

When the subharmonic mode begins to grow and overwhelms the Rossby wave, these induction components decrease.  Instead, the interaction of the subharmonic with itself  becomes larger than its weak-field counterpart \anna{(solid green line in figure~\ref{fig:dmd_forces})}, indicating enhancement of induction.  Since the Rossby wave component in the magnetic field is weak in this regime, it appears that the  triads of $(\mathbf{u}^{SW}, \mathbf{B}^{SW}, \mathbf{B}^{M})$ are preferred by the flow over $(\mathbf{u}^{SW}, \mathbf{B}^{SW}, \mathbf{B}^{RW})$. Consistently with that, the dominant contributions to induction come from the interaction of the subharmonic in the flow with the mean magnetic field, which can be again interpreted as $\alpha$-effect, and the subharmonic in the field with the enhanced zonal flow mode $u^M$ \anna{(dash-dotted line in figure~\ref{fig:dmd_forces}a)}, a proxi for $\Omega$-effect of toroidal field generation by differential rotation. Both these interactions transfer energy to the subharmonic wave and reinforce it, and $\alpha-\Omega$ dynamo becomes the principal dynamo mechanism during the transient subharmonic state. 

The expansion for the Lorentz force $\mathbf{F}_L$ is derived similarly to equation~\eqref{eq:force_decomp} as a sum of products of different magnetic components in~\eqref{eq:field_decomp} with induced by them currents $\nabla \times \mathbf{B}^{M}$, $\nabla \times \mathbf{B}^{RW}$, $\nabla \times \mathbf{B}^{SW}$.  Figure~\ref{fig:dmd_forces}(b) shows the volume integral  of the  $\theta$-component of the Lorentz force, 
    \begin{equation}\label{eq:lorentz_force_int}
   |F_{L, \theta}|^V = \left[ (1/V)\int_V F_{L,\theta}^2 dV \right]^{1/2}.
\end{equation}
It indicates transition of the dominant feedback on the flow from the currents induced by the Rossby wave to those induced by its subharmonic, leading to a 5-times increase in rms force amplitude. Similarly to the induction term, large contributions to Lorentz force are given by interaction of electric currents induced by the magnetic part of the Rossby wave or its subharmonic with the same magnetic mode, e.g., $(\nabla \times \mathbf{B}^{RW}) \times \mathbf{B}^{RW}$, and with the mean magnetic field - e.g., $(\nabla \times \mathbf{B}^{SW}) \times \mathbf{B}^{M}$. A considerable contribution to Lorentz force arises from interaction of currents induced by the large-scale current, $\nabla \times \mathbf{B}^{M}$, and the mean field $\mathbf{B}^{M}$. Similar behaviour is found in the radial and azimuthal components of the Lorentz force. \anna{In $F_{L,\phi}$,} the interaction of the form $(\nabla \times \mathbf{B}^{SW}) \times \mathbf{B}^{SW}$ is the strongest, indicating potential mechanism of zonal flow enhancement  $\mathbf{u}_{M}$ in the transient regime. As for cross-modal interactions between the Rossby wave and the subharmonic, such as $\mathbf{u}^{SW} \times \mathbf{B}^{RW}$, $(\nabla \times \mathbf{B}^{RW})\times \mathbf{B}^{SW}$, these are active only during transitions between different unstable flow states, \anna{around}  $t\approx 1.4$ and $t\approx 2.0$.

\subsection{Taylorization and Taylor-Proudman constraint}\label{sec:force_balance}
\begin{figure}
    \centering
\includegraphics[width=\textwidth]{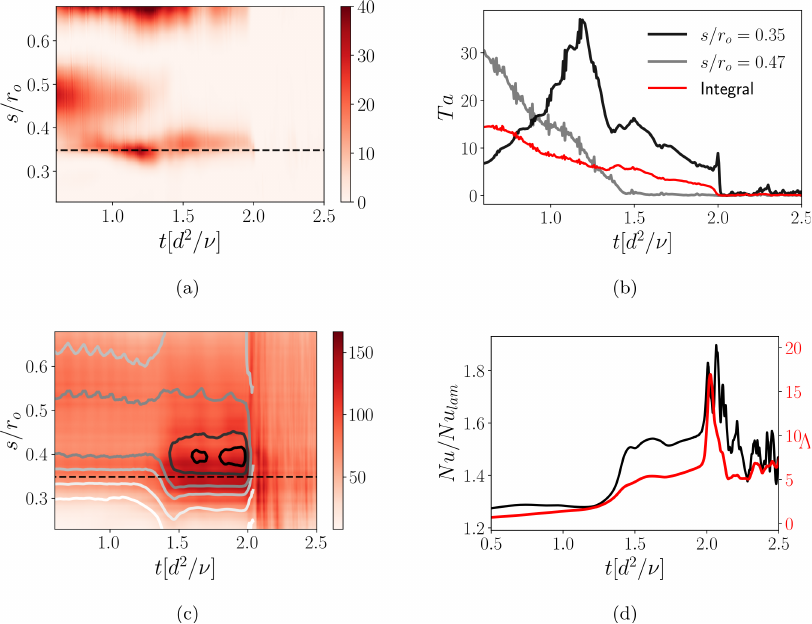}

    \caption{(a) Taylorization measure $Ta$ as a function of cylindrical radius $s$ and time. Dashed line indicates the tangent cylinder, $s/r_o = 0.35$. (b) $Ta$ at $s/r_o = 0.35$ and $s/r_o = 0.47$, together with the integral over the radius $s$ (in red). (c) In red: the distribution of velocity gradients $\int_{C(s)} |\partial u /\partial z | dz d\phi$, normalized by the surface area of $C(s)$. Grey contours denote the normalized integral for velocity, $\int_{C(s)} |u| dz d\phi$, at increasing levels of $[10,25,50,70,100,118]$.  (d)  Nusselt number $Nu$ at the bottom of the convective zone, normalized with its laminar value $Nu_{lam}$ (in black), and Elsasser number \anna{$\Lambda$} (in red). $Ra=180$, initial conditions at $Ra=160$.}
    \label{fig:taylicity}
\end{figure}

During the subharmonic state, both flow and magnetic field reconfigure so that their amplitude exhibits a maximum around the tangent cylinder \anna{(figure~\ref{fig:dmd_modes}e)}, and strong gradients develop at high latitudes in its vicinity. To investigate the role of the tangent cylinder in the transition, we introduce two characteristic measures of force balance in the system, integrated over the curved cylindrical surface $C(s)$ of radius $s$, coaxial with the rotation axis. The first one, ``Taylicity"
\begin{equation}\label{eqn:taylorization}
    T_{a} =  \frac{\left(s \int_{C(s)} \left[(\nabla \times \mathbf{B}) \times \mathbf{B} \right] \mathbf{e_\phi} dz d\phi \right)^2}{ <B^2>^2},
\end{equation}
was proposed by \citet{li2018taylor} to measure how well Taylor constraint and thus magnetostrophic \anna{(MAC)} force balance is satisfied (see section~\ref{sec:intro} for more details). \anna{This quantity measures strength of the azimuthal component of the Lorentz force and is proportional to magnetic tension. In figure~\ref{fig:taylicity}(a) we plot it} at different cylindrical radii during transition. In the weak dynamo state,  $T_a$ is maximal in two regions: near the equator, where both magnetic field and stresses are very small, and around $s=0.47 r_0$. As the flow undergoes the subharmonic instability, this maximum moves towards the tangent cylinder, denoted by the dashed line. \anna{The azimuthal magnetic force peaks at the tangent cylinder at about $t\approx 1.3$, during build-up of the subharmonic state}.
 These results suggest that the subharmonic dynamo state is  not in the magnetostrophic force balance.  When the subharmonic state itself becomes unstable and chaotic strong-field dynamo sets in, $T_a$ approaches zero both at the tangent cylinder and integrally, indicating that the dynamo is entering magnetostrophic force balance everywhere in the domain (figure~\ref{fig:taylicity}).

According to Taylor-Proudman theorem for hydrodynamic flows, axial velocity gradients should be zero in the limit of strong rotation, a constraint that is broken when magnetic tension enters the dominant force balance. Thus the second measure,
\begin{equation}\label{eqn:grad_u}
   T_u (t,s) =  2 d \frac{\int_{C(s)} | \partial \mathbf{u} / \partial z| dz d\phi }{\int_{C(s)} |\mathbf{u}| dz d\phi},
\end{equation} 
compares characteristic velocity gradients along rotation axis $z$ to the typical value of velocity as a function of time and cylindrical radius.  Figure~\ref{fig:taylicity}(c) shows the colormap of the integrated velocity gradients $\int_{C(s)} | \partial \mathbf{u} / \partial z| dz d\phi$ as a function of time and $s$, together with the contours of velocity measure $ \int_{C(s)} |\mathbf{u}| dz d\phi$, both divided by the area  $2 \pi d s $ of each cylindrical surface $C(s)$. It appears that velocity increases in the vicinity of the tangent cylinder as the subharmonic mode develops. This result holds even when the mean zonal flow $V_\phi$ is subtracted, indicating the topological change in the flow structure accompanying transition. At the same time, velocity gradients develop a maximum at the tangent cylinder while the outer flow regions remain less affected. The magnitude of velocity is comparable to the magnitude of velocity gradients (see contour levels in figure~\ref{fig:taylicity}c), resulting in nearly constant value of $T_u \approx 6$ throughout the entire transition. Thus, enhanced velocity gradients along $z$-axis, signalling less vertically constrained flow, do not necessarily indicate that the MAC force balance is achieved.

 Finally, figure~\ref{fig:taylicity}(d) illustrates the effect of these topological changes on two other diagnostic quantities, Nusselt number $Nu$ and Elsasser number $\Lambda$,
 \begin{equation}\label{eq:NuLam}
     Nu = \anna{ \frac{W_{heat}}{W_{cond}}}, \qquad  \Lambda = \frac{B^2_{rms}}{2\Omega\rho \mu \eta}.
 \end{equation}
 $Nu$, defined as a ratio between total heat flux \anna{$W_{heat}$} and conductive heat flux \anna{$W_{cond}$}, measures heat transport efficiency at the outer boundary $r_o$; $\Lambda$ measures the ratio of Lorentz to Coriolis force. \anna{While the subharmonic mode promotes at the same time 50\% more efficient mode of convection and five time stronger fields, these characteristics reach their peak values only when  the subharmonic itself becomes unstable}.  After that,  they relax back to the subharmonic levels in the strong-field state.

\section{Run-away transition to chaos and strong dynamo}\label{sec:runaway}

\begin{figure}
    \centering
 \includegraphics[width=\textwidth]{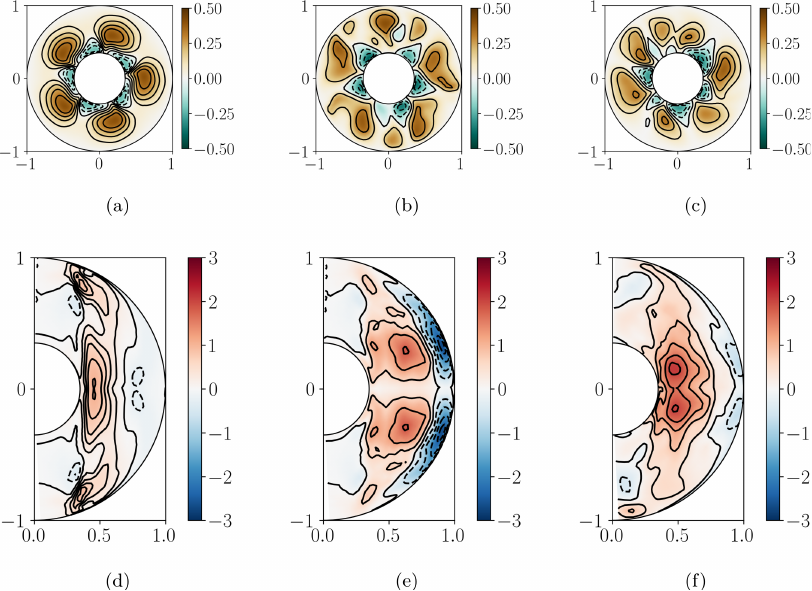}      
    \caption{Instantaneous snapshots of (a-c) equatorial slices of temperature $T$; (d-f) axisymmetric component of the field $B_\theta$. (a,d) Subharmonic transient state before the runaway energy growth ($1.5<t<2.0$); (b,c) flow state at the peak of energy in figure~\ref{fig:w2s_dns}(b), $t\approx 2.0$; (c,f) strong-field multimodal dynamo in chaotic state ($t>2.0$). $Ra=180$, initial conditions at $Ra=160$.}
    \label{fig:trans_runaway}
\end{figure}

The transient subharmonic dynamo ends up in a rapid increase of  kinetic and magnetic energy at about $t=2$ (figure~\ref{fig:w2s_dns}b).  The contribution to the mean temperature profile and thus heat transfer also peak at this time (figure~\ref{fig:dmd_coeffs}b).  On the other hand, the subharmonic mode decays in the magnetic field, together with the mean magnetic modes and the zonal flow (figure~\ref{fig:dmd_coeffs}a,b).  This run-away process is much faster then the growth of the subharmonic mode, and is followed by its breakdown into a turbulent strong-field dynamo state. The resulting axisymmetric dipolar field in the strong-field regime (figure~\ref{fig:w2s_dns}c, right) resembles this of the subharmonic mean-field mode (figure~\ref{fig:dmd_modes}c), although the regions of the strongest magnetic field are no longer located at the tangent cylinder. 

\begin{figure}
    \centering
\includegraphics[width=\textwidth]{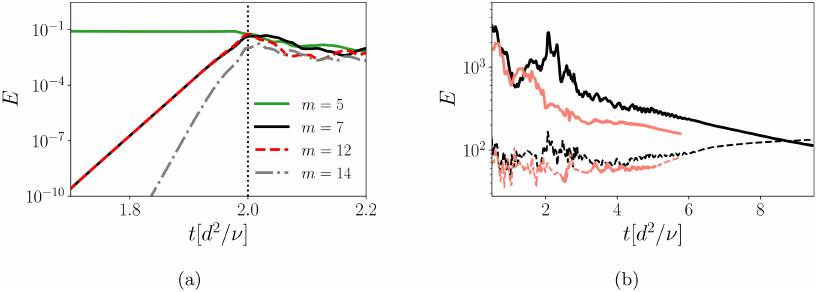}      
    \caption{(a) Evolution of energy in different azimuthal structures of magnetic field during and after the subharmonic state. Black solid line, $m=7$, green solid line, $m=5$, dashed, $m=12$, dashdotted, $m=14$. Vertical dotted line represets transition to chaos as in figure~\ref{fig:w2s_dns}(b). $Ra=170$, initial conditions at $Ra=160$. (b) Decay of the strong dynamo and inverse transition to the weak regime, $Ra=140$ (in black, initial condition at $Ra=160$) and $Ra=130$ (initial condition at $Ra=150$, in red). }
    \label{fig:chaos_m}
\end{figure}

To understand better this transition, we follow the evolution of the dynamo energy spectra in time, here using the spherical harmonic representation employed in PaRoDy \cite{dormy1998mhd}.  During the weak and the transient regimes, the energy is contained primarily in the azimuthal mode of $m=5$, with a relatively simple flow structure  (see temperature snapshots in figure~\ref{fig:trans_runaway}a). In the transient regime, however, the subharmonic state allows to transfer energy to the rest of azimuthal wavenumbers. Figure~\ref{fig:chaos_m}(a) illustrates this process on the azimuthal structures of $m=7$ and $m=12$,  which exhibit the fastest exponential growth of $\sigma_r=33$ from nearly zero initial energy.  In fact, all the azimuthal wave numbers are excited, some of them with the same growth rate of $m=7$, and some of them, like $m=14$, growing two times faster, $\sigma_r \approx 65$. Their growth seems to be determined by secondary quadratic interactions of the first group, and their energy remains smaller at all times than the energy of the triad $m=(5,7,12)$. When $m=7$ and $m=12$ enter in equipartition with the subharmonic mode at $t\approx2$, the total flow energy peaks (figure~\ref{fig:w2s_dns}b), indicating a triadic resonance of the three structures. Only then the presence of $m=7$ and $m=12$ becomes apparent in the dynamo snapshots, which show convective rolls visibly splitting in half and moving outwards in the equatorial plane (figure~\ref{fig:trans_runaway}b). Small flow scales imply larger dissipation, and so this flow configuration is highly unstable. Immediately after $t=2$, the energy of $m=12$ decays and the flow relaxes to the turbulent state still dominated by subharmonic-like structure, with contribution of $m=7$  and lower-order azimuthal modes (figure~\ref{fig:trans_runaway}c). This is when the energy fluctuations develop in figures~\ref{fig:w2s_dns}(b),~\ref{fig:chaos_m}(a), and the dynamo becomes \anna{strong-field}.

Topological changes in the flow result in respective adjustment of the magnetic field. Figures~\ref{fig:trans_runaway}(d-f) shows the latitudinal component of axisymmetric magnetic field $\langle B_\theta \rangle_\phi$, for illustration of dipolar field evolution. During the subharmonic transient,  the maximum intensity of the field is observed at the tangent cylinder, and is  correlated with the strong prograde zonal flow \anna{(figure~\ref{fig:dmd_modes}d)}. When convective structures are expelled outwards during triadic resonance, magnetic flux is also expelled from the tangent cylinder towards the outer sphere (figure~\ref{fig:trans_runaway}e).  The zonal flow (not shown here for brevity) is disrupted and considerably weakens in amplitude; the prograde jet bends towards the equator at high latitudes, constraining the location of the retrograde flow to the equatorial area. As $m=12$ mode decays, magnetic structures \anna{migrate} to the bulk of the domain (figure~\ref{fig:trans_runaway}f), and the zonal flow reconfigures to a state similar to the subharmonic, although a slight bent towards the equator remains in the zonal flow (figure~\ref{fig:w2s_dns}d, right).

\section{Discussion}\label{sec:discussion}

Subcritical transition to strong dipolar magnetic field configuration in planetary dynamo models produces topological changes in magnetic and velocity fields and leads to departure from the weak-field regime, dominated by convective Rossby waves. Using  Dynamic Mode Decomposition, we have identified that these changes are facilitated by a subharmonic mode of the Rossby wave, which drives a strong prograde zonal flow near the tangent cylinder. In this unstable flow state, the mode corresponding to convective Rossby wave remains weak, and the local maxima of dipolar  axisymmetric magnetic field shift from the area inside the tangent cylinder to  the tangent cylinder itself. A possible explanation for this phenomenon is that in the weak-field regime magnetic flux is generated by convective columns in the bulk and is transported by them inside the tangent cylinder. With strong shear at the tangent cylinder, this process is inhibited because the azimuthal jet transforms inward radial magnetic flux into toroidal, and prevents the accumulation of the net magnetic flux near the axis of rotation. These topological changes modify dominant dynamo mechanisms, as seen from the contribution of the mean, Rossby and subharmonic DMD modes to the generation of induction and Lorentz forces in the system. In the weak dynamo regime, the dominant  nonlinearities are of the type $(RW, RW, M)$ and $(RW, M, RW)$ in all induction components, corresponding to the driving of mean field through fluctuation interactions ($\alpha$-effect) and perturbation of the field by the Rossby wave.  In the subharmonic flow regime, the mean zonal flow enters the dominant nonlinear interactions with the subharmonic wave of the type $(M, SW, SW)$ and $(SW,SW,M)$ (see figure~\ref{fig:dmd_forces}). This suggests that the dynamo generating mechanism changes from $\alpha^2$ to $\alpha-\Omega$ as the dynamo enters the subharmonic regime. Furthermore, as the flow transitions to the turbulent strong-field regime, its topology remains similar to that of the subharmonic state (figure~\ref{fig:trans_runaway}d,f). Although in this state turbulent fluctuations are able to advect magnetic structures, generated in the bulk of the flow, inside the tangent cylinder, local magnetic maxima remain weak there, indicating that the shear is still dynamically active (figures~\ref{fig:w2s_dns}c,d). Despite developing turbulence, the axisymmetric dynamo configuration remains predominantly dipolar throughout the entire transition.

The dynamo is able to evolve around the subharmonic state for a considerable amount of time, independent of $Ra$; however, this state is relatively regular and does not explain the transition to the turbulent strong-field dynamo. Our results show that the strong-field state is reached after the runaway exponential growth of the flow energy in azimuthal modes  $m=7$ and $m=12$, entering triadic resonance with the quasi-steady subharmonic mode of $m=5$. The resonance is accompanied by expulsion of the dipolar magnetic field away from the tangent cylinder \anna{(figure~\ref{fig:trans_runaway}e)}. After the decay of the small-scale mode \anna{with} $m=12$, the dynamo relaxes to a distorted chaotic state with  flow and field structure similar to that of the subharmonic. Only then  surface-averaged azimuthal magnetic stresses decrease by the order of magnitude; thus, the dominant magnetostrophic balance is achieved in the strong-field turbulent state but not in the subharmonic regime, where magnetic stresses are large at the tangent cylinder  (figure~\ref{fig:taylicity}a,b). Despite several growing azimuthal wavenumbers (figure~\ref{fig:chaos_m}a), the energy distribution as a function of spherical degree $l$ (not shown here for brevity) remains similar through the whole transition. Therefore, the dynamically important direction of transition to the strong-field dynamo is azimuthal angle $\phi$, not latitude $\theta$.

To understand better the origin of resonance modes, we compared their growth rates with the growth rates of convective hydrodynamically unstable modes in table~\ref{tab:growth_rates}. The latter were computed with small perturbations as initial conditions and nonlinear feedback terms switched off in equations~\eqref{eq:NStInd}. The dominant wavenumbers of convectively unstable modes are smaller-scale structures of $m=7$, $8$, and $9$, comparable to the ones responsible for triadic resonance. Although the hydrodynamic instability develops faster than the resonance or secondary modes, their growth rates are comparable and  are distinctly different from much slower growth rates of the kinematic weak dynamo and the subharmonic mode itself. This suggests that the subharmonic state with its slightly increased magnetic field facilitates the growth  of hydrodynamically unstable convective modes that are otherwise inhibited by magnetic field, favouring large-scale structures.  The differences in the growth rates could arise due to the fundamental changes in the base flow state, with a flatter mean temperature profile and non-zero mean magnetic field. Similar lowering of convective threshold in the presence of moderate  magnetic fields was observed recently in simulations of magnetoconvection  \cite{mason2022magnetoconvection}.

\begin{table}
\centering
\begin{tabular}{|c|c |c |c |c| c|} 
 \hline
 $Ra$&Convection & Weak dynamo & Subharmonic  & Resonance modes   & Secondary modes \\ [0.5ex] 
 \hline
 160& 190 & 0.436 &  &   &\\ 
 170 & 208 & 0.434 & 2.174 & 33.243 & 65.139 \\ 
 \hline
\end{tabular}
\caption{Estimated growth rates of several instabilities in the dynamo flow. From the left to the right: hydrodynamic convective instability (dominant azimuthal wavenumbers $m=7$, $8$, $9$), kinematic growth of the weak magnetic field due to it, growth rate of the subharmonic mode ($m=5$), growth of dominant resonance modes  $m=7$ and $m=12$, and their secondary nonlinear quadratic interactions (i.e., $m=14$) at $Re=170$.}
\label{tab:growth_rates}
\end{table}

During the inverse transition from strong to weak-field dynamo \anna{regimes}, when $Ra$ is decreased beyond stability of the strong-field branch (figure~\ref{fig:w2s_dns}a), the flow does not revisit \anna{unstable} subharmonic or resonant flow states (figure~\ref{fig:chaos_m}b). Instead, the inverse transition takes place through a gradual relaxation of chaos in previously excited azimuthal modes, while the dominant convective mode grows as the only one able to extract  energy from the mean temperature profile.  The existence of the two unstable states is constrained by the region of bi-stability in figure~\ref{fig:w2s_dns}(a), where the two locally attracting weak- and strong-field states serve as boundaries. The chaotic strong-field solution decays when $Ra$ is low and  outside of this region, so the only remaining attractors for these runs are either the weak-field dynamo or purely non-magnetic convective states at even lower $Ra$.  \anna{The subharmonic or resonant unstable states} could be found exactly using adjoint techniques \cite{mannix2022systematic} or edge-tracking bisection method \cite{guseva2017azimuthal}.

Our simulations employed no-slip boundary conditions, appropriate for terrestial planets with rigid mantle. Simulations with free-slip or mixed boundary conditions, more suitable in gas and ice giant planets, allow for a different type of bi-stability between strong-field dipolar and weak-field multipolar states \cite{sasaki2011weak}. In the absence of viscous Ekman layers near rigid boundaries, stronger zonal flows are able to develop, saturated only  through bulk viscosity and magnetic stresses, if present.  These large-scale flows stretch magnetic field lines, inducing a stronger $\Omega$-effect and promoting multipolar, sometimes oscillatory magnetic fields \cite{schrinner2012dipole}.  Although such transitions also can take place in simulations with rigid boundaries, they require strong turbulent convection \cite{petitdemange2018systematic}. In this work, both weak and strong dynamo states develop very close to the onset of convection, and are predominantly dipolar, so the zonal flow is enhanced in the strong-field turbulent state. The distribution of zonal flows changes from three-layered to two-layered during the transition from the weak to strong regimes (figure~\ref{fig:w2s_dns}d), contrary to those observed in dipolar-multipolar transitions \cite{schrinner2012dipole}. Thus, subcritical transitions provide a potential mechanisms of simultaneous enhancement of magnetic field and zonal flows when overall convective efficiency is weak - for example, hampered by opposing chemical gradients during planetary formation~\cite{leconte2012new}.  \anna{Initial conditions for the magnetic field during such transitions can be important for developing planetary or stellar magnetic fields and can define their magnitude \cite{morin2011weak}.}

\section{Conclusions}

\anna{With an innovative data-driven method, we highlighted}  nonlinear interactions between physical effects that play a key role in the generation of  planetary magnetic fields. Our 3D convective dynamo simulations in spherical geometry can model these links in a certain numerically accessible parameter space. We were particularly interested in transitions that give rise to strong dipolar fields like those observed for most planets \cite{schubert2011planetary, christensen2011geodynamo,roberts2013genesis}. Such strong dipolar fields imply a hierarchy of forces in the dynamo zone with predominance of magnetic, buoyancy and Coriolis force \cite{aubert2017spherical,menu2020magnetic,teed2023solenoidal}, see \citet{tobias2021turbulent} and \citet{moffatt2019self} for a review. \anna{This force balance is essential for driving torsional or magnetic Rossby waves and thus for detection of planetary magnetic  dynamics, such as temporal variations of the Earth's magnetic field on short timescales \cite{gillet2010fast,buffett2016evidence,hori2022waves}.}

Nonlinearity of convective dynamos allows bi-stability of dynamo solutions: a weak field branch where the magnetic field is not
strong enough to enter into the leading order force balance in the momentum equation, and a strong field branch where the field enters into the balance, at least at certain scales. The transitions between the two, occurring with varying intensity of convection and magnetic induction, can be either gradual (supercritical) or abrupt (subcritical). In this work, we analysed the subcritical transition using Dynamic Mode Decomposition, identifying  dynamically active flow scales.  Our analysis shows that its physical origin  is related to sufficiently strong magnetic field being able to induce subharmonic instability of the dynamo on the weak-field branch. The subharmonic dynamo state, itself unstable, is  followed by a triadic resonance of convective modes, leading to turbulence and thus to a more efficient mode of convection and dynamo. Subharmonic instability is accompanied by topological changes in the flow: development of strong two-layered shear in the vicinity of the tangent cylinder, and \anna{build-up} of magnetic flux outside the tangent cylinder. This distribution of the flow and the  field remains dynamically active later, during the turbulent strong-field state.  Furthermore,  we obtained modal structures spanning the whole dynamical landscape of this flow;  subsequently, we aim to construct data-driven  reduced-order models of this transition using  Sparse Identification of Nonlinear Dynamics \cite{brunton2016discovering} or Galerkin methods \cite{noack1994low}.

In the future work, we will focus on such transitions in planetary dynamos at lower Ekman and higher Rayleigh numbers, where the weak-field branch corresponds to multipolar dynamos~\cite{petitdemange2018systematic}, and investigate whether the transition scenario shares similar features with this case. The physical processes described here give rise to a turbulent dynamo dominated by a magnetostrophic balance at large scales, and may also play a role in the formation of dipole-dominated magnetic fields during magnetic reversals. 
Furthermore, subcritical convective dynamos like those reported here can help to understand evolution of planetary magnetic fields, e.g. to identify dynamo extinction scenarios in planets with remnant magnetism like Mars whose cores sufficiently cooled down~\cite{schubert2011planetary}, or in non-magnetic planets like Venus with \anna{core-mantle cooling potentially insufficient} to drive convection~\cite{smrekar2018venus}.

\section*{Open Research Section}

The simulations in this work were performed using code PaRoDy \cite{dormy1998mhd,aubert2008magnetic}. Their parameters and initial conditions, together with resulting numerical data and Python scripts supporting this work are available at the repository Zenodo with CC BY 4.0 license \cite{guseva2024dataset}.

\section*{Acknowledgements
}

This work has received funding from the European Union’s Horizon 2020 research and innovation programme under the Marie Skłodowska-Curie grant no. 890847 and European Research Council (ERC) grant no. D5S-DLV-786780. It was partly supported by funding from the French program 'PROMETHEE' managed by Agence Nationale de la Recherche and from "Visiting Fellowship" managed by Observatoire de Paris-PSL. This study used the HPC resources of MesoPSL financed by the Région île-de-France and the project EquipMeso (reference ANR-10-EQPX-29-01) of the program Investissements d'Avenir, supervised by the Agence Nationale pour la Recherche.

\bibliographystyle{IEEEtranN}
\bibliography{cnrs}

\end{document}